# Mechanism Design with Execution Uncertainty

Ryan Porter [†]          Amir Ronen [†‡]          Yoav Shoham [†]          Moshe Tennenholtz [†]

[†]Computer Science Department                    [‡]ICSI
Stanford University                               UC Berkeley
{rwporter,amirr,shoham,tennenholtz}@robotics.stanford.edu     amirr@icsi.berkeley.edu

## Abstract

We introduce the notion of *fault tolerant mechanism design*, which extends the standard game theoretic framework of mechanism design to allow for uncertainty about execution. Specifically, we define the problem of task allocation in which the private information of the agents is not only their costs to attempt the tasks, but also their probabilities of failure. For several different instances of this setting we present technical results, including positive ones in the form of mechanisms that are incentive compatible, individually rational and efficient, and negative ones in the form of impossibility theorems.

## 1 INTRODUCTION

Recent years have seen much activity at the interface of computer science and game theory, in particular in the area of Mechanism Design, or MD (e.g. (Parkes & Ungar 2000; Boutilier, Shoham, & Wellman 1997; Shoham & Tennenholtz 2001; Nisan & Ronen 2001)). A sub-area of game theory, MD is the science of crafting protocols for self-interested agents, and as such is natural fodder for computer science in general and AI in particular. The uniqueness of the MD perspective is that it concentrates on protocols for non-cooperative agents. Indeed, traditional game theoretic work on MD focuses uniquely on the incentive aspects of the protocols.

A promising application of MD to AI is the problem of task allocation among self-interested agents (see e.g. (Rosenschein & Zlotkin 1994)). When only the execution costs are taken into account, the task allocation problem allows standard mechanism design solutions.

However, this setting does not take into consideration the possibility that agents might fail to complete their assigned tasks. When this possibility is added to the framework, existing results cease to apply. The goal of this paper is to investigate robustness to failures in the game theoretic framework in which each agent is rational and self-motivated. Specifically, we consider the design of protocols for agents which have not only private cost functions, but also privately-known probabilities of failure.

What criteria should such protocols meet? Traditional MD has a standard set of criteria for successful outcomes, namely social efficiency (maximizing the sum of the agents' utilities), individual rationality (positive utility for all participants), and incentive compatibility (incentives for agents to reveal their private information). Fault Tolerant Mechanism Design (FTMD) strives to satisfy these same goals; the key difference is that the agents have richer private information (namely probability of failure, in addition to cost). As we will see, this extension presents novel challenges.

It is important to distinguish between different possible types of failure. The focus of this paper is on failures that occur when agents make a full effort to complete their assigned tasks, but may fail. A more nefarious situation would be one in which agents may also fail deliberately when it is rational to do so. While we do not formally consider this possibility, we will revisit it at the end of the paper to explain why our results hold in this case as well. Finally, one can consider the case in which there exist irrational agents whose actions (for example, intentional failures) are counter to their best interests. This is the most difficult type of failure to handle, because the presence of such agents can affect the strategy of rational agents, in addition to directly affecting the outcome. We leave this case to future work.

It is helpful to consider a concrete example. Consider a network of links which are owned by selfish agents

---

[1]This work was supported in part by DARPA grant F30602-00-2-0598.

(e.g. airline companies), and two distinguished nodes $S$ and $T$ in it. We allow multiple links between nodes so that more than one agent can provide the same service (but only one agent can be selected to do so). When an object is routed through a link, the owning agent incurs some cost. In addition, the agent may *fail* (according to some probability) to pass the object across the link (e.g., the object is lost in transit, or not delivered by a strict deadline). The costs and probabilities are *privately* known to their owners. Our goal is to design a mechanism (protocol) that will ensure that objects will be sent from $S$ to $T$ across the network in the most reliable and cost-effective way possible.

To demonstrate the challenges encountered when facing such problems, consider even the simple case in which the network consists of only parallel links between $S$ and $T$, and costs are all zero. A naïve protocol would ask each agent for their probability, choose the most reliable agent (according to the declarations) and pay her a fixed, positive amount if she succeeds, and zero otherwise. Of course, in this case each agent will report a probability of one in order to selfishly maximize her *own* expected profit.

In this paper we study progressively more complex task-allocation problems. The first problem that we study is one in which there is only one task. We use this setting both to show why standard MD solutions are not applicable and to present our basic technique in the form of a novel mechanism. After extending this technique to handle the case of multiple tasks without dependencies among them, we move to the general case of dependent tasks. Here, we prove an impossibility result when we demand incentive compatibility in dominant strategies, and present a mechanism that solves in equilibrium the case of dependent tasks. Finally, we discuss the use of cost verification to significantly improve the revenue properties of the center.

## 2  RELATED WORK

The work presented in this paper integrates techniques of economic mechanism design (an introduction to MD can be found in (Mas-Collel, Whinston, & Green 1995, chapter 23)) with studies of fault tolerant problem solving in computer science and AI.

In particular, the technique used in our mechanism is similar to that of the Generalized Vickrey Auction (GVA) (Vickrey 1961; Clarke 1971; Groves 1973) in that it aligns the utility of the agents with the overall welfare. This similarity is almost unavoidable, as this alignment is the only known general principle for solving mechanism design problems. However, because we allow for the possibility of failures, we will need to change the GVA in a significant way in order for our mechanism to achieve this alignment.

Because we have added probabilities to our setting, our mechanisms may seem to be related to the Expected Externality Mechanism (or d'AGVA) (d'Aspremont & Gerard-Varet 1979), but there are key differences. In the setting of d'AGVA, the types of the agents are drawn (independently) from a distribution which is assumed to be common knowledge among the participants. The two key differences in our setting are that no such common knowledge assumption is made and that the solution concepts which we guarantee are stronger than that of d'AGVA.

A recent paper (Eliaz 2002) also considers failures in MD, but solves a different problem. This work assumes that agents know the types of all other rational agents and also limits the failures that can occur by bounding the number of irrational agents.

Finally, the design of protocols which are robust to failures has a long tradition in computer science (for a survey, see (Linial 1994)). Work in this area, however, almost always assumes a set of agents that are by and large cooperative and adhere to a central protocol, except for some subset of malicious agents who may do anything to disrupt the protocol. In MD settings, the participants fit neither of these classes, but are simply self-interested.

## 3  A BASIC MODEL

In this section we describe our basic model and notation, which will be modified later to handle specific settings.

In a FTMD problem, we have a set of $t$ tasks $\tau = \{1, \ldots, t\}$ and a set $N = \{1, \ldots, n\}$ of self-interested agents to which the tasks can be assigned. We also have a center $M$ who assigns tasks to agents and pays them for their work. The center and the agents will collectively be called the participants.

Each agent $i$ has, for each task $j$, a *probability* $p_{ij} \in [0, 1]$ of successfully completing task $j$, and a nonnegative *cost* $c_{ij} \in \Re^+$ of attempting the task. We assume that the cost of attempting a task does not depend on the success of the attempt. We use $p_i = (p_{i1}, \ldots, p_{it})$ for the set of all probabilities for agent $i$, and use $p = (p_1, \ldots, p_n)$ to represent the set of probability vectors for all agents. We use corresponding notation for $c_i$ and $c$. The pair $\theta_i = (p_i, c_i)$ is called the agent's *type* and is *privately* known to the agent. Each agent is assigned a set $A_i$ of tasks, and her cost to attempt the set is: $c_i(A_i) = \sum_{j \in A_i} c_{ij}$. We define $\theta = (\theta_1, \ldots, \theta_n)$ as the vector of types for all agents.

We use a completion vector $\mu \in \{0, 1\}^t$ to denote which tasks have been completed. The function $V : \{0,1\}^t \rightarrow \Re^+$ defines the center's nonnegative valuation for each possible completion vector. For now, we assume that the center has a non-combinatorial valuation for a set of tasks. That is, the value of a set of tasks is the sum of the values for the individual tasks. We also assume that $V(\mu) \geq 0$ for all $\mu$ and that $V(0,\ldots,0) = 0$.

An assignment vector $A = (A_1,\ldots,A_n)$ and a vector of agent probabilities $p$ together induce a probability distribution over the completion vector which we denote by $[\mu|A,p]$. Given an assignment $A$, a type vector $\theta$ and a completion vector $\mu$, we define the *welfare* of the participants as $W(A,c,\mu) = V(\mu) - \sum_i c_i(A_i)$. We define the *expected welfare* as $\bar{W}(A,c,p) = E_{[\mu|A,p]}[W(A,c,\mu)]$. The goal of the center is to design a mechanism (protocol) that maximizes this *expected welfare*.

We assume that each task can be assigned only once. The center does not have to allocate all the tasks. For notational convenience we assume that all the non-allocated tasks are assigned to a dummy agent 0 which for each task has zero probability of success and zero cost to attempt.

When an agent $i$ is assigned a set $A_i$ of tasks, and is paid $R_i$, her *utility* equals $u_i = R_i - c_i(A_i)$. Since our setting is stochastic by nature, an agent can do no better than to maximize her *expected utility*, $\bar{u}_i$, calculated before any task is attempted. This term thus depends on the true probabilities of success of the agents, as explained below.

Throughout the paper we shall use the following vector notations: The subscript $-i$ on a vector denotes that the term for agent $i$ has been omitted from the vector. For example, $p_{-i} = (p_1,\ldots,p_{i-1},p_{i+1},\ldots,p_n)$. The omitted term can be combined with such a vector by using the following notation: $p = (p_i, p_{-i})$. We denote by $\mu_i$ the completion vector for agent $i$ (i.e. we have 1 for each task accomplished by agent $i$ and 0 for each one either failed by her or not assigned to her). The definitions for $\mu_{-i}$ and $(\mu_i, \mu_{-i})$ follow similarly. Sometimes we will use $\mu_i$ in place of $p_i$. Since both vectors are of the same form, a 0 or 1 for task $t_j$ in $\mu_i$ becomes the probability of successfully completing $t_j$.

### 3.1 Mechanisms

A mechanism is a protocol that decides how to assign the tasks to the agents and how much each agent is paid. The simplest type of mechanisms are ones in which the agents are simply required to report their types. (Of course they may lie!) The revelation principle (see e.g. (Mas-Collel, Whinston, & Green 1995, p. 871)) tells us that we can, w.l.o.g., restrict ourselves to such mechanisms.

We denote by the vector $\hat{\theta}$ the types declared by the agents . A mechanism is thus defined by a pair $g = (A(\hat{\theta}), R(\hat{\theta},\mu))$ such that:

- $A(\hat{\theta}) = (A_1(\hat{\theta}),\ldots,A_n(\hat{\theta}))$ is an assignment function. It takes a declaration vector and returns an assignment of the tasks to the agents.

- $R(\hat{\theta}) = (R_1(\hat{\theta},\mu),\ldots,R_n(\hat{\theta},\mu))$ is the payment function.

In our motivating example, a type $\theta_i$ would correspond to agent $i$'s costs and probabilities of success on each of her edges.

In our protocol, the center first asks each agent to declare her type. We call an agent *truthful* if she reveals her true type to the center. Based on these declarations the center first computes the assignment $A(\hat{\theta})$. Then, the agents execute their tasks. Finally, the center pays the agents. Note that these payments depend on the set of tasks which were accomplished. We assume that the agents always attempt each task to which they are assigned. In our discussion section, we explain why this is a valid assumption.

In the above protocol, the utility of agent $i$ is: $u_i(c_i,\hat{\theta}_i,\hat{\theta}_{-i},\mu) = R_i(\hat{\theta},\mu) - c_i(A_i(\hat{\theta}))$, and her expected utility is: $\bar{u}_i(c_i,\hat{\theta}_i,\hat{\theta}_{-i},p) = E_{[\mu|A(\hat{\theta}),p]}[u_i(c_i,\hat{\theta}_i,\hat{\theta}_{-i},\mu)]$.

The main difference between mechanism design problems and the usual algorithmic problems is that the participating agents may *manipulate* the given protocol if it is *beneficial* for them to do so. We therefore need to design protocols that fulfill our objectives even though the agents behave selfishly. We thus require our mechanism to satisfy the following standard properties:

*Individual rationality (IR)* holds when truthful agents are guaranteed to have non-negative expected utility. Formally, for all $i$, $\theta$ and $\hat{\theta}_{-i}$: $\bar{u}_i(c_i,\theta_i,\hat{\theta}_{-i},p) \geq 0$.

*Incentive compatibility (IC)* holds when it is a dominant strategy for each agent to declare her type truthfully. Formally, this condition holds, when for all $i$, $\theta$, $\theta_i'$, and $\hat{\theta}_{-i}$: $\bar{u}_i(c_i,\theta_i,\hat{\theta}_{-i},p) \geq \bar{u}_i(c_i,\theta_i',\hat{\theta}_{-i},p)$. This means that the expected utility of the agent (conditional on her own probability of success) is maximized when the agent reports her true type.

A mechanism is called *socially efficient (SE)* if the chosen assignment maximizes the expected welfare $\bar{W}$. The fact that $\bar{W}$ depends on the true types of the agents underscores the importance of IC, which allows the center to correctly assume that $\hat{\theta} = \theta$.

*Individual rationality for the center (CR)* holds if the center's utility $u_M = V(\mu) - \sum_i R_i(.)$ is *always* nonnegative. CR is an extension of the standard mechanism design requirement of weak budget balance to account for the center's utility for outcomes.

A final goal is *no free riders (NFR)*, which holds if all agents not assigned any task have a revenue of zero.

# 4　SINGLE TASK SETTING

We will start with the special case of a single task to show our basic technique to handle the possibility of failures in MD. For expositional purposes, we will analyze two restricted settings (the first restricts probabilities of success to be one, and the second restricts costs to be zero), before formally proving properties about our mechanism in the full single task setting.

Because there is only one task, we can simplify the notation. We let $c_i$ and $p_i$ denote $c_{i1}$ and $p_{i1}$, respectively. Similarly, we let $V = V((1))$, which is the value that the center assigns to the completion of the task. For each mechanism, we will use the index [1] to denote the agent selected to attempt the task (e.g., $p_{[1]}$ denotes the selected agent's probability of success). The subscript [2] then refers to the agent who would be selected as the service provider if agent [1] had not participated.

## 4.1　CASE 1: ONLY COSTS

When we do not allow for failures (that is, $\forall i\ p_i = 1$), the goal of social efficiency reduces to assigning the task the lowest-cost agent. This simplified problem can be solved using a second-price auction (which is a specific case of GVA). Each agent declares a cost, the task is assigned to the agent with the lowest cost, and that agent is paid the second-lowest submitted cost.

## 4.2　CASE 2: ONLY FAILURES

We now reduce the problem in a different way, by assuming all costs to be zero ($\forall i\ c_i = 0$). In this case, the main goal is to allocate the task to the most reliable agent. Interestingly, we cannot use a straightforward application of the GVA for this case. Such a mechanism would ask each agent to declare a probability of success and assign the task to the agent with the highest declared probability. It would set the revenue function for all agents not assigned the task to be 0, while the service provider would be paid the amount by which her presence increases the welfare of the other agents and the center: $\hat{p}_{[1]}V - \hat{p}_{[2]}V$. Obviously, such a mechanism is not incentive compatible, because the payment to the service provider depends on her own declared type! Since there are no costs, each agent would have a dominant strategy to declare her probability of success as one. [2]

Thus, we need to fundamentally alter our payment rule so that it depends on the outcome of the attempt, and not solely on the declared types, as it does in the GVA. The key difference in our setting that forces this change is the fact that the true type of an agent now directly affects the outcome, whereas in the standard MD setting the type of an agent only affects her preferences over outcomes. We accomplish our goals by replacing $\hat{p}_{[1]}$ with an indicator function that is 1 if the task was completed, and 0 otherwise. The payment rule for the service provider is now $V - \hat{p}_{[2]}V$ if she succeeds and $-\hat{p}_{[2]}V$ if she fails. Just as in the previous setting, the service provider is the only agent who has positive utility for attempting the task with the corresponding payment rule. The expected utility for agent $i$ would be $V \cdot (p_i \cdot (1 - \hat{p}_{[2]}) - (1 - p_i) \cdot \hat{p}_{[2]})$. This expression is positive for the agent iff $p_i > \hat{p}_{[2]}$, which is only true for the service provider.

## 4.3　CASE 3: COSTS AND FAILURES

We now consider the case of one task with both costs and failures.

We introduce the following definition that we will use throughout the paper: Given an agent $i$ we denote by $\bar{W}^*_{-i}(\hat{c}_{-i}, \hat{p}_{-i})$ the optimal (declared) expected welfare when tasks cannot be allocated to agent $i$. In the single task case it is $max_{j \neq i}(\hat{p}_j \cdot V - \hat{c}_j)$. Now we can define the mechanism.

**Single Task Mechanism:**

**Assignment** The mechanism chooses an agent $i \in \{0, \ldots, n\}$ that maximizes the (declared) expected welfare $\hat{W} = \hat{p}_i \cdot V - \hat{c}_i$.

**Payment** The payment to all agents not assigned a task is always zero. The payment to the "winner" $i$ is defined as follows:

$$R_i = \begin{cases} V - \bar{W}^*_{-i}(\hat{c}_{-i}, \hat{p}_{-i}) & \text{If } i \text{ succeeds} \\ -\bar{W}^*_{-i}(\hat{c}_{-i}, \hat{p}_{-i}) & \text{If } i \text{ fails} \end{cases}$$

Using $p_{[2]}$ and $c_{[2]}$ to denote the probability and the cost of the "second best" agent, the payment to agent $i$ when she succeeds is $(V - p_{[2]} \cdot V + c_{[2]})$ and when she fails is $(-p_{[2]} \cdot V + c_{[2]})$. Note that $\bar{W}^*$ is never negative

---

[2]In fact, this would be true for any payment rule for which an agent's payment is always nonnegative, which is the reason why we require our goals (such as IC and IR) to be satisfied for the expected utility of the agent, and not for ex post utility.

because the center will never make an assignment that yields an expected loss for the system.

For example, suppose we have three agents with the types listed in Table 1. Let $V$ be 210. If the agents are truthful, then the winner is agent 3. If agent 3 did not exist, the optimal expected welfare would be $\bar{W}^*_{-3} = 210 - 100 = 110$, because the task would be assigned to agent 2. The payment for agent 3 is therefore $210 - 110 = 100$ if she succeeds and $-110$ if she fails. Agent 3's own costs are 60, and thus her expected utility is $(100 - 60) \cdot 0.9 + (-110 - 60) \cdot 0.1 = 19$.

| Agent | $c_i$ | $p_i$ |
|-------|-------|-------|
| 1     | 30    | 0.5   |
| 2     | 100   | 1.0   |
| 3     | 60    | 0.9   |

Table 1: A Single Task Example

Before we prove the properties of this mechanism, let us introduce two definitions that we shall use throughout that paper. Given an agent $i$, we define the welfare of the other participants $W_{-i}(A, c_{-i}, \mu) = V(\mu) - \sum_{j \neq i} c_j(A_j)$. Note that $W_{-i}(A, c_{-i}, \mu) = W(A, c_i, \mu) + c_i(A_i)$. We define the expected welfare for the other participants as $\bar{W}_{-i}(A, c_{-i}, (p_{-i}, \mu_i)) = E_{[\mu_{-i}|A, p_{-i}]}[W_{-i}(A, c_{-i}, (\mu_i, \mu_{-i}))]$. This is the expected welfare of all the other participants (including the center) when the allocation is $A$ and agent $i$ has completed exactly the set of tasks defined by $\mu_i$.

It is not difficult to see that the payment $R_i$ of each agent $i$ equals $\bar{W}_{-i}(A, \hat{c}_{-i}, (\hat{p}_{-i}, \mu_i)) - \bar{W}^*_{-i}(\hat{c}_{-i}, \hat{p}_{-i})$. Her expected utility is therefore $\bar{u}_i = -c_i(A_i) + E_{[\mu_i|A_i, p_i]}\bar{W}_{-i}(A, \hat{c}_{-i}, (\hat{p}_{-i}, \mu_i)) - \bar{W}^*_{-i}(\hat{c}_{-i}, \hat{p}_{-i})$. Since the distribution $[\mu|A, (p_i, \hat{p}_{-i})]$ equals $[\mu_i|A_i, p_i] \cdot [\mu_{-i}|A, p_{-i}]$, we get that $R_i = \bar{W}(A, (c_i, \hat{c}_{-i}), (p_i, \hat{p}_{-i})) - \bar{W}^*_{-i}(\hat{c}_{-i}, \hat{p}_{-i})$. This means that each agent gets paid her *contribution* to the expected welfare of the other participants.

**Theorem 1** *The Single Task mechanism satisfies IR, IC, CR, SE, and NFR.*

**Proof:**

We will prove each property separately.

1. **Individual Rationality**

   We need to prove that the expected utility of a truthful agent is always non negative. When agent $i$ is truthful her expected utility is $\bar{u}_i = \bar{W}(A((\theta_i, \hat{\theta}_{-i})), (c_i, \hat{c}_{-i}), (p_i, \hat{p}_{-i})) - \bar{W}^*_{-i}(\hat{c}_{-i}, \hat{p}_{-i})$. By the optimality of $A(.)$, the second term quantifies the optimal welfare that can be obtained when the types of the other agents are

   $\hat{\theta}_{-i}$ and $i$ does not exist. Similarly, the first term quantifies the optimal welfare when the types of the other agents are $\hat{\theta}_{-i}$ and the type of agent $i$ is the true one $\theta_i$. Since $i$ can only improve the total welfare, we proved our claim.

2. **Incentive Compatibility**

   We need to prove that the expected utility of each agent $i$ is maximized when she is truthful. Let $\theta_{-i}$ denote the declarations of the other agents. As before, when the agent is truthful her utility is $\bar{u}_i = \bar{W}(A((\theta_i, \hat{\theta}_{-i})), (c_i, \hat{c}_{-i}), (p_i, \hat{p}_{-i})) - \bar{W}^*_{-i}(\hat{c}_{-i}, \hat{p}_{-i})$. Consider the case where the agent reports another type $\theta'_i$. This results in an assignment $A'$. The utility of agent $i$ in this case is $\bar{u}'_i = \bar{W}(A', (c_i, \hat{c}_{-i}), (p_i, \hat{p}_{-i})) - \bar{W}^*_{-i}(\hat{c}_{-i}, \hat{p}_{-i})$

   Assume by contradiction that $\bar{u}'_i > \bar{u}_i$. This means that $\bar{W}(A', (c_i, \hat{c}_{-i}), (p_i, \hat{p}_{-i})) > \bar{W}(A((\theta_i, \hat{\theta}_{-i})), (c_i, \hat{c}_{-i}), (p_i, \hat{p}_{-i}))$. However this contradicts the optimality of $A((\theta_i, \hat{\theta}_{-i}))$.

3. **Individual Rationality for the Center** Let agent $i$ be the winner. We need to show that the utility for the center is always non negative. There are two cases. When agent $i$ succeeds we have $u_M = V - (V - \bar{W}^*_{-i}(\hat{c}_{-i}, \hat{p}_{-i})) = \bar{W}^*_{-i}(\hat{c}_{-i}, \hat{p}_{-i}) \geq 0$. When $i$ fails, the value is zero and thus $u_M = \bar{W}^*_{-i}(\hat{c}_{-i}, \hat{p}_{-i}) \geq 0$.

4. **Social Efficiency** Immediate from IC and the definition of $A(.)$.

5. **No Free Riders** Immediate from the definition of the payment rule. ■

# 5    MULTIPLE TASKS

We now return to the original setting presented in this paper, consisting of $t$ tasks for which the center has a non-combinatorial valuation (that is, the value for a set of tasks is equal to the sum of the values for the individual tasks). Because the setting disallows any interaction between tasks, we can construct a mechanism that satisfies all of our goals by generalizing the Single Task Mechanism.

**Multiple Task Mechanism:**

**Assignment** The chosen assignment $A$ maximizes the (declared) expected welfare $\bar{W}(A, \hat{c}, \hat{p}) = E_{[\mu|A, \hat{p}]}[W(A, \hat{c}, \mu)]$.

**Payment** The payment to each agent $i$ is defined according to her completion vector: $R_i = \bar{W}_{-i}(A, \hat{c}_{-i}, (\hat{p}_{-i}, \mu_i)) - \bar{W}^*_{-i}(\hat{c}_{-i}, \hat{p}_{-i})$

In other words, each agent is paid according to her contribution to the welfare of the other participants.

**Proposition 2** *The Multiple Task mechanism satisfies IC, IR, SE, CR, and NFR.*

The proof is similar to the single task case and is omitted. Note that the theorem holds even when the cost functions and probabilities of success have a combinatorial nature.

### 5.1 COMBINATORIAL $V$

We now consider the setting in which the center's valuation $V(\cdot)$ can be any monotone function of the tasks. Unfortunately, in this setting, it is impossible to satisfy all our goals simultaneously.

**Theorem 3** *When $V$ is combinatorial, there does not exist a mechanism that satisfies IC, IR, CR, and SE for any $n \geq 2$ and $t \geq 2$.*

Intuitively it is enough to consider the following case which no mechanism is able to solve. There are two tasks, each of which can only be completed by one agent (and, this one agent is different for the two tasks). The center only has a positive value (call it $x$) for both tasks being completed. Since both agents add a value of $x$ to the system, they can each extract close to $x$ from the center, causing the center to pay double for the utility of $x$ he will gain from the completion of the task. Due to space constraints, we omit the formal proof of this theorem.

However, despite the possibility of failures we can maintain the desired properties other than CR using the same mechanism as before.

**Theorem 4** *The Multiple Task mechanism satisfies IC, IR, SE, and NFR, even when $V$ is combinatorial.*

Again, we omit the proof. Intuitively, IC, IR, and NFR are not affected by a combinatorial $V$ because they are only properties of the agents, and SE still follows from IC and the definition of $A(\cdot)$.

### 5.2 DEPENDENCIES

We now return to the case of non-combinatorial valuation $V(\cdot)$, and analyze a different extension: dependencies between the tasks.

Consider our motivating example of a network of flights. A natural example of a task dependency would be an object that could not be carried over the edge $(b, c)$ before being carried over $(a, b)$.

Formally, we say that a task $j$ is *dependent* on a set $s$ of tasks if $j$ cannot be attempted unless all tasks in $s$

were successfully finished. We assume that there are no dependency cycles. The tasks now are executed according to a topological order. Note that if a task cannot be attempted, the agent assigned that task does not incur the costs of attempting it. [3]

However, the presence of dependencies, just like the presence of a combinatorial $V$, makes it impossible to satisfy IC, IR, CR, and SE.

**Theorem 5** *When dependencies exist between tasks, there does not exist a mechanism that satisfies IC, IR, CR, and SE for any $n \geq 2$ and $t \geq 2$.*

**Proof:** Proof by induction. We first show that a mechanism cannot satisfy IC, IR, CR, and SE for the base case of $n = t = 2$. The inductive step then shows that increasing either $n$ or $t$ cannot alter this impossibility result.

**Base Case:** We prove the base case by contradiction. Assume that there exists a mechanism that satisfies IC, IR, CR, and SE. This implies that it satisfies these properties for all possible instances, where we define an instance as a particular set of agent types and declarations, task dependencies, and a $V$ function. We will use 3 possible instances in order to derive properties that must hold in the mechanism, but lead to a contradiction. The constants in these instances are that task 2 is dependent on task 1 and that the center has value of 5 for task 2 being completed, but no value for the completion of task 1 in isolation. The five types that we will use, $\theta_1$, $\theta_1'$, $\theta_1''$, $\theta_2$, and $\theta_2'$, are defined in Table 2 (the final type, $\theta_e$, will be used later in the inductive step).

| | | | | |
|---|---|---|---|---|
| $\theta_1:$ | $p_{11} = 1$ | $c_{11} = 2$ | $p_{12} = 1$ | $c_{12} = 1$ |
| $\theta_1':$ | $p_{11}' = 1$ | $c_{11}' = 2$ | $p_{12}' = 0$ | $c_{12}' = 0$ |
| $\theta_1'':$ | $p_{11}'' = 1$ | $c_{11}'' = 0$ | $p_{12}'' = 1$ | $c_{12}'' = 4$ |
| $\theta_2:$ | $p_{21} = 0$ | $c_{21} = 1$ | $p_{22} = 0$ | $c_{12} = 0$ |
| $\theta_2':$ | $p_{21}' = 1$ | $c_{21}' = 1$ | $p_{22}' = 0$ | $c_{22}' = 0$ |
| $\theta_e:$ | $p_{e1} = 0$ | $c_{e1} = 1$ | $p_{e2} = 0$ | $c_{e2} = 1$ |

Table 2: Agent Types for Theorem 5.

*Instance 1:* The true types are $\theta_1$ and $\theta_2$, and the declared types are $\theta_1$ and $\theta_2'$. To satisfy SE, task 1 is assigned to agent 2, and task 2 to agent 1. That is, $A_1(\theta_1, \theta_2') = (0, 1)$ and $A_2(\theta_1, \theta_2') = (1, 0)$. Since agent 2's true type is $\theta_2$, she will fail on task 1, preventing task 2 from being attempted. Thus, $\mu = (0, 0)$ with probability 1. The expected utility for agent 1 is then:

$$\bar{u}_1(c_1, \theta_1, \theta_2', p) = R_1((\theta_1, \theta_2'), (0, 0))$$

---

[3] This is the reason why the current setting is not a special case of the combinatorial $V$ setting where the valuation of a set of tasks is the valuation of the subset for which the prerequisites are met.

*Instance 2:* The true types are $\theta_1'$ and $\theta_2$, and the declared types are $\theta_1$, and $\theta_2'$. Thus, the only difference from instance 1 is agent 1's true type which is insignificant, because agent 1 never gets to attempt a task. Thus, we have a similar expected utility function:

$$\bar{u}_1(c_1', \theta_1, \theta_2', (p_1', p_2)) = R_1((\theta_1, \theta_2'), (0, 0))$$

*Instance 3:* The true types are $\theta_1'$ and $\theta_2$, and the declared types are $\theta_1'$, and $\theta_2'$. Now we have also changed agent 1's declared type to $\theta_1'$. Both tasks will be allocated to the null agent: $A_1(\theta_1', \theta_2') = A_2(\theta_1', \theta_2') = (0, 0)$. Therefore, $\mu = (0, 0)$ still holds with probability 1, and we get the following equations:

$$\bar{u}_1(c_1', \theta_1', \theta_2', (p_1', p_2)) = R_1((\theta_1', \theta_2'), (0, 0))$$

$$\bar{u}_2(c_2, \theta_2', \theta_1', (p_1', p_2)) = R_2((\theta_1', \theta_2'), (0, 0))$$

If $R_2((\theta_1', \theta_2'), (0, 0)) < 0$, then IR would be violated if $\theta_2'$ were indeed the true type of agent 2, because the assignment function would be the same. Since the center thus receives no payment from agent 2, and it never gains any utility from completed tasks, the CR condition requires that $R_1((\theta_1', \theta_2'), (0, 0)) \leq 0$. Thus, $\bar{u}_1(c_1', \theta_1', \theta_2', (p_1', p_2)) \leq 0$.

Notice that if agent 1 lied in this instance and declared her type to be $\theta_1$, then we are in instance 2. So, to preserve IC, agent 1 must not have incentive to make this false declaration. $\bar{u}_1(c_1', \theta_1, \theta_2', (p_1', p_2)) = R_1((\theta_1, \theta_2'), (0, 0)) \leq \bar{u}_1(c_1', \theta_1', \theta_2', (p_1', p_2)) \leq 0$.

*Instance 1:* Now we return to instance 1. Having shown that $R_1((\theta_1, \theta_2'), (0, 0)) \leq 0$, we know that when agent 1 declares truthfully in this instance, her expected utility will be: $\bar{u}_1(c_1', \theta_1, \theta_2', p) \leq 0$.

We will now show that agent 1 must have a positive expected utility if she falsely declares $\theta_1''$. In this case, both tasks are assigned to agent 1. That is, $A_1(\theta_1'', \theta_2') = (1, 1)$. We know that $R_2((\theta_1'', \theta_2'), (1, 1)) \geq 4$ by IR for agent 1, because if $\theta_1''$ were agent 1's true type, then both tasks would be completed and agent 1 would incur a cost of 4.

We now know that if agent 1 falsely declares $\theta_1''$ in instance 1: $\bar{u}_1(c_1', \theta_1'', \theta_2', p) = R_1((\theta_1'', \theta_2'), (1, 1)) - (c_{11} + c_{12}) \geq 4 - 3 \geq 1$. Thus, agent 1 has incentive to falsely declare $\theta_1''$ in instance 1, violating IC. Thus, we have reached a contradiction and completed the proof of the base case.

**Inductive Step:** We now prove the inductive step, which consists of two parts: incrementing $n$ and incrementing $t$. In each case, the inductive hypothesis is that no mechanism satisfies IC, IR, CR, and SE for $n = x$ and $t = y$, where $x, y \geq 2$.

*Part 1:* For the first case, we must show that no mechanism exists that satisfies IC, IR, CR, and SE

for $n = x + 1$ and $t = y$, which we will prove by contradiction. Assume that such a mechanism does exist. There is a one-to-one mapping from every instance in which $n = x$ and $t = y$ to an instance that only differs in the addition of an "extra" agent who truthfully declares her type $\theta_e$. Since such an instance satisfies $n = x + 1$ and $t = y$, our mechanism must satisfy IC, IR, CR, and SE for this instance. Because of SE, this mechanism can never assign the task to the extra agent. Because of IR, this mechanism can never receive a positive payment from the extra agent. Since in each instance the only effect that the extra agent can have on the mechanism is to receive a payment from the center, we can transform this mechanism into one which satisfies IC, IR, CR, and SE for all instances where $n = x$ and $t = y$ by simply removing the revenue function and assignment function for the extra agent, contradicting the inductive hypothesis.

*Part 2:* For the second case, we need to show that no mechanism can satisfy IC, IR, CR, and SE for $n = x$ and $t = y+1$. We use a similar proof by contradiction, starting from the assumption that such a mechanism does exist. There is a one-to-one mapping from every instance in which $n = x$ and $t = y$ to an instance of $n = x$ and $t = y+1$ through the addition of an "extra" task $t_e$ that is not involved in any dependencies and for which the center has no value for its completion. By SE, if a satisfying mechanism exists, then there exists a satisfying mechanism that always assigns this task to the dummy agent (recall that this is equivalent to not assigning the task). We can transform this mechanism into one which satisfies our goals for $n = x$ and $t = y$ by simply removing the assignment of $t_e$ to the dummy agent. Once again, we have contradicted the inductive hypothesis, and the proof is complete. ∎

Interestingly, by slightly altering our mechanism we can solve the problem in an equilibrium.

**Equilibrium Multiple Task Mechanism:**

**Assignment** The chosen assignment $A$ maximizes the (declared) expected welfare $\bar{W}(A, \hat{c}, \hat{p}) = E_{[\mu|A, \hat{p}]}[W(A, \hat{c}, \mu)]$.

**Payment** The payment to each agent $i$ is defined according to her completion vector: $R_i = \bar{W}_{-i}(A, \hat{c}_{-i}, \mu) - \bar{W}_{-i}^*(\hat{c}_{-i}, \hat{p}_{-i})$

The only difference from the Multiple Task Mechanism is that the first term of the payment rule uses the *actual* completion vector, instead of the distribution induced by the declaration of the other agents. As a result, our properties are satisfied only as an equilibrium: if all agents declare truthfully, then no agent has incentive to deviate to a different declaration. While there is no formal name for this type of equilibrium, it is similar in spirit to a Nash equilibrium, but technically different because there is no common knowledge of the game (since privately-known types affect the utility of other agents). It is also similar to a Bayes-Nash equilibrium, but much stronger because it holds regardless of agent beliefs about the prior distributions for the types of the other agents. We define Equilibrium IC to hold if truth-telling is such an equilibrium. Equilibrium IR and SE are defined similarly.

**Theorem 6** *The Equilibrium Multiple Task Mechanism satisfies NFR, Equilibrium IC, Equilibrium IR, and Equilibrium SE, even when dependencies exist.*

# 6   COST VERIFICATION

A practical drawback of our mechanisms is that the payments (or fines) may be very large, especially when service provider is far more efficient than the other agents. Also, since CR is not satisfied in our most general settings, the designer has to take a risk.

Previous work (Nisan & Ronen 2001) has stressed the importance of ex post verification. It showed that when the designer can verify the costs and the actions of the agents *after* the work was done, the power of the designer increases dramatically. All of our previous constructions have corresponding versions that use verification. The main advantage of these mechanisms is that the payments can be *normalized* by any linear function, thus making the potential losses more reasonable for both the agents and the center. Due to space constraints we omit these constructions.

# 7   DISCUSSION & FUTURE WORK

In this paper we studied task allocation problems in which agents may fail to complete their assigned tasks. For the settings we considered (single task, multiple tasks with combinatorial properties, and multiple tasks with dependencies) we provided either a mechanism that satisfies our goals or an impossibility result.

It is worth pointing out that all of the results in this paper hold when we expand the set of possible failures to include rational, intentional failures, which occur when an agent increases her utility by not attempting an assigned task (and thus not incurring the corresponding cost). Modelling this possibility would complicate our model without changing any of our results. Intuitively, our positive results continue to hold because the payment rule aligns an agent's utility with the welfare of the system. If failing to attempt some subset of the assigned tasks would increase the welfare, then these tasks would not have been assigned to any agent. Ob-viously, all impossibility results would still hold when we expand the set of possible actions for the agents.

Many interesting directions stem from this work. Two possibilities are retrying tasks after a failure or allowing multiple agents to attempt the same task in parallel. The computation of our allocation and payment rules presents non-trivial algorithmic problems. Also, the payment properties for the center may be further investigated, especially in settings where CR must be sacrificed to satisfy our other goals.

Finally, we believe that the most important future work will be to consider a wider range of possible failures, and to discover new mechanisms to overcome them. In particular, we would like to explore the case in which agents may fail maliciously or irrationally. For this case, even developing a reasonable model of the setting provides a major challenge.